# A multiscale generative model to understand disorder in domain boundaries

**Authors:** Jiadong Dan[1,2,8,*], Moaz Waqar[3,4,8], Ivan Erofeev[1,5], Kui Yao[3], John Wang[3,4], Stephen J. Pennycook[6,7], N. Duane Loh[1,2,5,*]

**Affiliations:**

[1]NUS Centre for Bioimaging Sciences, National University of Singapore, 14 Science Drive 4, 117557, Singapore.

[2]Department of Biological Sciences, National University of Singapore, 16 Science Drive 4, Singapore, 117558, Singapore.

[3]Institute of Materials Research and Engineering (IMRE), A*STAR (Agency for Science, Technology and Research), Singapore 138634, Singapore.

[4]Department of Materials Science and Engineering, National University of Singapore, Singapore, 117574, Singapore.

[5]Department of Physics, National University of Singapore, 2 Science Drive 3, Singapore, 117551, Singapore.

[6]Department of Materials Science and Engineering, University of Tennessee, Knoxville, TN, USA.

[7]School of Physical Sciences and CAS Key Laboratory of Vacuum Sciences, University of Chinese Academy of Sciences, Beijing, China.

[8]These authors contributed equally: Jiadong Dan, Moaz Waqar.

[*]Correspondence to Jiadong Dan, N. Duane Loh.

**Abstract**

A continuing challenge in atomic resolution microscopy is to identify significant structural motifs and their assembly rules in synthesized materials with limited observations[1]. Here we propose and validate a simple and effective hybrid generative model capable of predicting unseen domain boundaries in a potassium sodium niobate thin film from only a small number of observations, without expensive first-principles calculation. Our results demonstrate that complicated domain boundary structures can arise from simple interpretable local rules, played out probabilistically. We also found new significant tileable boundary motifs and evidence that our system creates domain boundaries with the highest entropy. More broadly, our work shows that simple yet interpretable machine learning models can help us describe and understand the nature and origin of disorder in complex materials.

**Introduction**

Condensed matter systems are often described by simple and finite motifs (or structural units), and complex structures can arise from simple assembly rules for these motifs. In a seminal paper, Frank and Kasper invoked the idea that complex alloy structures could be viewed as sphere packings through geometrical constraints[2]. Inspired by the polyhedral arrangements in these alloy systems, Bernal used "canonical polyhedra" to describe monoatomic liquids, which is consistent with the experimental data[3]. In metallic glasses, the "stereochemical" model offers a similar picture that bonding units, which are almost identical to their crystalline analogs, interconnect to form amorphous alloys[4]. Throngs of these simple units often self-organize to form configurations that are unexpectedly complicated and elicit intensive theoretical and experimental structure-property studies[5–11]. Using aberration-corrected scanning transmission electron microscopy (STEM), we can now observe these fundamental units (*i.e.*, recurrent arrangements of atoms called *structural motifs*) at the atomic scale. In the past few decades, this basic idea of structural motifs has been useful for modeling structures of grain boundaries[12–16], crystals[17], quasicrystals[18–21], amorphous solids[22,23], colloids[24,25], and supramolecular lattices[26]. Despite the success of structural motifs and their correlation to materials' function and properties, the inter-motif relationships in real physical systems remain largely unexplored, especially in disordered systems where there is a wide variety of motif-motif realizations.

Inter-motif relationships build up the microscopic structure, which in turn will dictate materials' macroscopic properties. For instance, regular tetrahedra, when connected in different ways, can form dense ordered structures[19], which might offer tangible connections between local motifs and the structure of bulk condensed matter[5]. The ways these tetrahedra pack into three-dimensional structures even affect the packing fraction. The interconnection of structural motifs can also play a role in the disorder-order transition, as seen in metallic glasses where the formation of "superclusters" can lead to the emergence of new order [4,27]. The relationship between structural motifs is often studied using methods such as common neighbor analysis[28] and clustering techniques[1], but these analyses are limited in their ability to extrapolate and understand the connection between motifs and disorder.

Motifs are expected to assemble with disorder into complex structures, hence maximizing their entropy. As these complex motifs grow in size, different constraints may start to dominate. This emergence of complex and consequential mesoscale and macroscale structures often leads to new simplifying descriptions, sometimes even new Physics, emerging[29,30].

One of the simplest ways that atomic disorder manifests in materials is at the grain interfaces (*i.e.*, interfaces between two regions of the same crystal species, including grain boundaries and domain boundaries). Structural motifs can form complicated configurations along grain boundaries[13,14] and, in certain cases, dictate macroscopic properties[31,32]. A translation domain boundary (TDB) is rather a simpler case of boundaries that separates two regions in a material whose crystal lattices are misaligned by a specific translation that is not a unit cell vector. Fig. 1a shows an exemplary high-angle annular dark field (HAADF) STEM image of a potassium sodium niobate (KNN) thin film, where such TDBs are overlaid with false colors. The domains on two sides of a TDB are displaced by half a unit cell in the [100] or [010] lattice directions. This dense network of TDBs was reported to be related to the giant effective piezoelectric coefficient in nonstoichiometric alkali niobate epitaxial thin films deposited by sputtering[33,34]. While experiments and simulations have provided insights into the coexistence of different structural motifs and the kinetics of grain boundary transitions[35–37], current approaches seem to presume that these interfaces are dominated by repeating structural motifs that are functionally important – but the statistical frequencies that these motifs occur are not quantified.

Despite their low dimensionality, the structural complexity and diversity of grain interfaces have yet to be satisfactorily modeled or understood. Thus far, it is unclear if grain interfaces are "completely random" or adhere to organizational rules at different length scales. Here we use alkali-deficient KNN, investigated with aberration-corrected STEM (Fig. 1a), as the model system to understand four fundamental questions about structural motifs along the boundaries separating translational domains (Fig. 1b). 1). Can domain boundaries be reduced to a finite number of local structural motifs (*i.e.*, reductionism)? 2). How do structural motifs assemble into longer, and more complex motifs? 3). What are the significant long-sequence motifs that emerge? 4). At what length scales do new constraints dominate these increasingly complex long-sequence motifs?

To model the complexity of motif sequences on boundary structures, it is instructive to note its resemblance to complex word sequences in languages. Simple and interpretable language models (LMs) [38] often reduce text into a set of unique words (*i.e.*, motifs), which "probabilistically assemble" to form complex phrases that are governed by grammatical rules (*i.e.*, longer motifs). These complex phrases, in turn, exist within a context and must conspire to form an intelligible narrative. Analogously, using $NbO_6$ octahedra as building blocks (Fig. 1c), here we seek recurrent local structural motifs in TDBs in Fig. 1d (*i.e.*, words), and measure the conditional probabilities that can probabilistically generate realistic longer structural motifs in Fig. 1e (*i.e.*, phrases) at intermediate length scales. These intermediate-length structural motifs then collude to form the boundaries that demarcate competing translational domains in Fig. 1f (*i.e.*, context).

## Results

### A workflow to represent motifs and extract their assembly rules

The enthalpic and entropic terms of a system's free energy may dominate at different length scales. Therefore, a simplistic local model is usually inadequate for describing the system at these different length scales – this is the case with TBDs in Fig. 1a. Hence, a combination of graph representations and a hybrid generative model allows us to quantitatively examine the hierarchy of structures from small to large length scales. Figure 2 shows the three components of the workflow to construct a hybrid generative model of complex and random TDBs: First, extract and represent structural motifs from the planar graph constructed from atomic resolution STEM images (Fig. 2, a to c). Second, regress the probabilistic Markov chain model to generate random, medium-length motif sequences (Fig. 2, c to e).

Last, create a scaffold of coarse-grained boundary segments that are replaced by medium-length motif sequences in the previous step (Fig. 2, e to f).

Our workflow begins by extracting polygonal motifs, which are particularly efficient for representing atomically-resolved samples. Fig. 2a shows a HAADF-STEM image of a KNN thin film plan view with a high density of TDBs. We identify centers of $NbO_6$ octahedra as feature points (blue dots) and optimize their positions using a maximum filter method and center-of-mass refinement[1]. These octahedral $NbO_6$ units are ordered to form the perovskite phase. Without loss of generality, we then implement Voronoi neighbor analysis (see Methods) to locate neighbors of each feature point (node), and edges are constructed thereafter for each pair of nodes, forming the planar graph in Fig. 2b. A region extraction algorithm[39] obtains polygons in the planar graph. Four-sided polygonal motifs belonging to the crystalline phase within the translation domains are eliminated. Only the TDBs are colorized in Fig. 2c. For simplicity, we define individual polygons here as 1-motif (*i.e.*, a single motif); Two connecting 1-motifs can combine to form a single 2-motif; Two 2-motifs can organize themselves to obtain a 3-motif configuration (Fig. S1). If this process continues, by definition, *k*-motifs are connected motif sequences of path length *k*. It should be noted that although our motifs are presented in two-dimensional space, they represent a collection of atomic columns in three-dimensional space.

Having extracted the polygonal 1-motifs (Fig. 2a-c), we determine the constrained probabilistic rules for generating 2- and 3-motifs. The ensemble of these domain boundaries in Fig. 2c can be abstracted as longer sequences composed of a finite number of 1-motifs. Triangular and hexagonal 1-motifs can combine to form thirty-two 2-motifs states (motif-pairs) $S = \{s_0, s_1, s_2, …, s_{31}\}$. This combination can be described by the transition probabilities of a generative Markov model illustrated as a chord diagram in Fig. 2d. (Details of the generative model will be discussed in sections below, the chord diagram here serves as a model summary in the workflow.) By iteratively calling this Markovian system in a stochastic manner, akin to a self-avoiding random walk comprising 1-motifs, we can generate an ensemble of simulated TDBs comprising *k* 1-motifs (Fig. 2e).

This generative Markovian model, which captures local motif-motif relationships, cannot extrapolate beyond the mesoscopic length scale: it does not enforce the size distributions of domains and the potential "interactions" between adjacent TDBs. Instead, we introduce a coarse-grained model of competing translation domains that is more efficient in enforcing these constraints that dominate at longer-length scales. This coarse-grained model produces featureless line segments as domain boundaries (solid black lines of the domains in Fig. 2f), which we replace with randomly sampled mesoscale *k*-motif segments generated by the Markovian system in Fig. 2e.

**Generating medium-range *k*-motifs from self-avoiding walks (SAW)**

We hypothesize that the intermediate features of TBDs are essentially entropically driven: they comprise quasi-random sequences of motifs. A memoryless Markov state model is a natural prescription of how to compose such a motif sequence as a self-avoiding walk (SAW). We now demonstrate how to construct our Markov state model from motifs. Figure 3 shows a summary of motifs, motif states (2-motifs), and the transition matrix (TM). From left to right, it shows a hierarchical structure of forming random complex boundary configurations from low-level constrained structural motifs. The TDB region is mainly composed of triangular and hexagonal motifs according to the statistics of identified 1-motif polygons (Fig. S2). The pentagon 1-motifs, which account for only less than 4%, mostly appear in frustrated and structurally dissimilar regions (Fig. S3) and are thus ignored in model construction. Six directional 1-motifs are identified in Fig. 3a using force-relaxed classification[1]: these comprise four orientations of triangular 1-motifs (top), and two orientations of hexagonal 1-motifs (bottom). However, directly using these six 1-motifs as Markov states (Fig. S4a) may generate sequences with unrealistically overlapping motifs (Fig. S4b). To address this issue, we have to consider how a 2-motif overlaps with the next 2-motif to form a 3-motif. This formulation requires at least 2-motifs as Markov states. Fig. 3b shows 32 such 2-motif Markov states, which have distinct combinations of constituent 1-motifs and their relative translation vectors (Fig. S5). These 32 Markov states can be divided into three groups based on Euclidean isometry (rigid transformation): 0-7, 8-23, and 24-31. Subsequently, the transition matrix $P_{ij}$ can be estimated from experimental data simply by counting occurrences of 3-motifs (Fig. 3c), and $P_{ij}$ is the conditional probability of 2-motif state $i$ overlaps with 2-motif state $j$ within each 3-motif, where $0 \leq i \leq 31$ and $0 \leq j \leq 31$ (see Methods).

The transition matrix estimated from experimental data is sparse and block-like, which indicates that the parameter space of the transition probability matrix can be further reduced by symmetry (SI Section I, Fig. S6, and Table S1). The sparsity stems from the fact that composing 2-motifs is not arbitrary and has to follow spatial constraints of motif-motif rules. For example, 2-motif state 0, which starts from a blue triangular 1-motif and ends with an orange triangular 1-motif, can only be compatible with either 2-motif state 3 or 17 to form a 3-motif since they begin with an orange triangular 1-motif that matches the same ending orange 1-motif in the previous motif state 0. This transition matrix is a compact and interpretable representation of the domain boundaries in this sample. Given a random initial 2-motif (denoted as αβ), the transition matrix can probabilistically extend it by 1-motif (γ) (*i.e.*, where the first αβ-motif to the βγ 2-motif). By repeating this process we can generate arbitrarily long motif sequences along domain boundaries that were not observed in experiments. This generation is equivalent to a SAW, illustrated in Movie S1. Examples of such SAWs are shown in Fig. 3d.

**Coarse-grained model of domain boundaries**

Whereas the generative Markov model efficiently encodes how low-level motifs assemble, we find it only captures motif correlations to medium-length scales. Moreover, this model underrepresents smaller domains that are enclosed by tight loops (Fig. S7 and Fig. S8). These small domains, we assume, are formed when one type of epitaxial translation domain is hemmed in by another type. It is computationally expensive to study the formation of these domains across a wide range of length scales using first principles. This high cost, however, can be substantially reduced with appropriate coarse-graining of these boundary segments.

The assembly of $NbO_6$ octahedra into extended boundaries is constrained at multiple length scales. At longer length scales, the creation of such boundaries can be coarse-grained as competitions between random nuclei [40] as shown in Fig. 4a. At shorter length scales, however, the boundaries between these domains are subjected to energy and entropy constraints that are captured by the transition matrix in Fig. 3.

In Fig. 4b and 4c, we phenomenologically combine constraints from these two length scales into a hybrid model capable of generating realistic but unseen TDBs spanning 2-3 orders of magnitude in length. Due to the translation symmetry of the domains, there are two equivalent types of nucleation sites shown in Fig. 4a: type I (blue) and type II (orange). A Voronoi tesselation of these nucleation sites coarse-grains their TDBs (Fig. 4b). Neighboring domains of the same type will merge (gray dashed edges in Fig. 4b), while those of different types will coarse-grain actual domain boundaries (black edges in Fig. 4b). The latter are then dressed with sequential structural motifs generated using the Markov model as shown in Fig. 4c (see Methods).

In our nucleation model, the coarse-grained shape of the TDBs can be controlled by the density of nucleation sites. These shapes are parameterized by the ratio between the total path length of TDB segments ($L$), and their corresponding end-to-end distance ($R$) as shown in Fig. 4d. This relationship between $L$ v.s. $R$ can be plotted for different nucleation densities as shown in Fig. 4e. By increasing the nucleation density, we observe the decrease of $R$ for large $L$, reflective of more convoluted segments within a certain field of view. Not only can our hybrid model generate arbitrarily long realistic samples, but it also forms smaller closed domains (*i.e.*, loops in Fig. 4b) that are experimentally observed (Figs. S7 and S8).

**Validating the hybrid model**

We use four different sets of measures to validate our hybrid model at different length scales. First, the steady-state distribution of 2-motifs, $\pi$, computed from the Markov transition matrix (blue points in Fig. 5a) is consistent with the 2-motif frequency directly calculated from experiments (orange circles in Fig. 5a). These two frequency distributions show plateaus for each of the three isometric 2-motifs, where the average frequencies on each plateau match those of the isometry-reduced TM (black dashed line in Fig. 5a; also see SI Section I and Fig. S6). It is important to note that the non-uniform steady-state distributions here indicate that certain 2-motifs are more energetically favored to form than others, even when the isometries are accounted for.

Second, we examine the radial distribution function $g(r)$ of the centers of 1-motifs. Fig. 5 (b and c) respectively show the $g(r)$ from experiments and synthetic samples drawn from our hybrid model. Our hybrid model produces $g(r)$ that match those from the experiment up to a medium-range order (MRO) of 20 Å.

Third, to validate our hybrid model beyond the MRO regime we need to consider longer boundary segments. Specifically, we examine the relationship between the end-to-end distance $R_n$ vs motif path length $N$ of each boundary segment. Fig. 5d shows that this relationship in our hybrid model (blue and red dashed lines), with two different nucleation densities ($\rho$), bounds that of the experiments (blue disks and orange crosses). Fig. 5d shows that segments generated from a solely self-avoiding walk (SAW; *i.e.*, without nucleation model in Fig. 4) and non-self-avoiding random walk (RW) respectively over- and under-estimate the $R_n$ vs $N$ relationship seen in experiments. Notably, the experimental $R_n$ vs $N$ curves show an absence of loops larger than $N=55$, which is likely due to limited data.

Finally, the domain boundaries appear fractal-like, which are ubiquitous in disordered phenomena in nature [41–44]. Fig. 5e shows that the Hausdorff box-counting (see Methods) fractal dimension of boundary segments from our hybrid model concurs with those from the experiment Fig. 1f.

These four validation measures give us confidence that our hybrid model reproduces many of the spatial features and correlations cloaked in the sample over a broad span of length scales.

**Significant long sequences of motifs**

Because our hybrid model is generative, it allows us to explore statistically likely boundary segments beyond those that were experimentally observed. The polarization of these boundary segments is expected to have important implications on its giant piezoelectric properties [34]. It is common practice to calculate, from first principles, the properties of boundary segments that can be periodically tiled [45].

Fig. S10 shows a large number of possible boundary segments that could occur between two translation domains. However, not all of these possibilities are equally likely.

To seek significant tileable long sequences of $k$-motifs, we ranked the frequency of all experimentally observed $k$-motifs where $2 \leq k \leq 6$, as shwon in Fig. 6. The three most frequently occurring tileable $k$-motifs are identified by colored dashed boxes in Fig. 6a; their corresponding ADF-STEM images are shown in Fig. 6b-d, which compare well with their multislice simulations in Fig. 6e-g given their atomic models Fig. 6h-j. The chemical information within these $k$-motifs was inferred from their scattering line profiles (see Supplementary Fig. S9). As an example, the two blue dashed boxes in Fig. 6a show how triangular 1-motifs can be tiled as a long chain, whose properties have been investigated [45].

Here, we found two significant tileable motifs in Figs. 6c and 6d that have not been studied previously. Notably, these two tileable motifs both contain hexagonal 1-motifs, which in turn host two K/Na atomic columns (green spheres in Figs. 6i and 6j) in close proximity. Unlike the already studied boundary in Fig. 6b, those in Figs. 6c and 6d have different gaps and orientations between the translation domains, and likely have different piezoelectric properties. By conducting analysis such as shown there, it would be possible to optimize film microstructure for improved properties.

We also observed repeated patterns of pentagonal motifs (Fig. S11), but these do not occur nearly as frequently as those from triangular/hexagonal motifs in Fig. 6.

**The system maximizes the configurational entropy of domain boundaries**

Highly entropic grain boundaries can stabilize the interfaces between ordered domains[46,47]. This stabilization occurs when the increase in configurational entropy dominates over the internal energy cost of forming longer grain boundaries.

Our hybrid model can explore how configurational entropy increases with boundary segment length $N$ by generating realistic but unseen boundary segments. This computation can be challenging, whether from first principles, or coarse-grained models, or only from experimental observations.

Here we examine a "microcanonical" configurational entropy, by restraining the internal energy of a boundary segment. The non-uniform 2-motif frequencies in Fig. 5a show that certain motif-motif juxtapositions are energetically less favorable than others. This also suggests that the internal energy of a boundary segment is dominated by its composition of 1-motifs and 2-motifs. Figs. 7a, b show two boundary segments that have the same composition of 1-motifs and 2-motifs (hence also same path length $N$) and also have the same end-to-end distance $R$. These constraints, we submit, qualify them to be in the same "microcanonical ensemble". Fig. 7c shows that its number of configurational microstates $\Omega(N, R) \propto N^\alpha$ (see Methods). Hence, its configurational entropy increases logarithmically with path length $N$.

Furthermore, we found evidence that the system tends to maximize its configurational entropy. As established above, the 2-motif frequencies in Fig. 5a restrain this "microcanonical ensemble". However, these 2-motif frequencies do not directly constrain the parameter $d$ in the system's reduced transition matrix (SI Section I and Fig. S6). Simply put, $d$ is unconstrained in this "microcanonical ensemble". Nevertheless, $d$ is inferred to be around 0.6 from experimental observations (Table S1). Why did the system tend towards this value for $d$? By varying $d$ from 0.1 to 0.9, Figs. 7c,d show that the boundary segments' configurational entropy is maximized in our generative model also when $d \approx 0.6$. In our model, $d$ determines the occurrence of a particular 3-motif (brown matrix element in Fig. S6) that appears only consequential to entropy but not internal energy. Overall, it seems plausible that our KNN sample has also tuned $d$ to maximize the configurational entropy of its TDBs. This maximization supports the idea that high-entropy boundaries have a stabilizing effect[46,47].

**Discussion**

The structural diversity and disorder in materials can appear bewildering. Our workflow can help us prioritize *where* to pay attention. Here we show how disorder along boundaries can be decomposed into statistical rules. Because these rules in our model are interpretable, it lets us see how they were tuned to maximize entropy. The combination of two interpretable models, self-avoiding random walker and random nucleation model, explains the structure and correlation of domain boundaries across a wide range of length scales (Fig. 1c-f). This multiscale hybrid model provided a more comprehensive exploration of the KNN sample. The fractal-like nature of domain boundaries in KNN allows a dense packing of structural motifs that can be another factor contributing to the enhanced piezoelectricity. These wide, fractal-like domain boundaries maximize configurational entropy to stabilize the domains they straddle.

The exemplary KNN sample studied here shows a giant piezoelectric coefficient of ~1900 picometer per volt, and previous studies attributed it to a local antisite model[33] and a planar fault model[34]. Yet, our analysis identifies new tileable long sequences that are statistically significant, and their functional importance may further be studied in the future which may provide new insights into the physical mechanisms of large piezoelectric response. Furthermore, our analyses indicate that certain configurations are more energetically favorable than others. The properties and stability of these new structures could be investigated with density functional theory (DFT) and/or molecular dynamics (MD) to obtain the temperature dependence of the transition matrix, and hence guide strategies for their controlled synthesis. Our work can be generally extended and applied to other important systems where grain boundaries play an essential role in controlling the physical properties of materials. For example, grain boundaries in metals and alloys show complex structures and their transformational kinetics effectively control grain growth at high temperatures[35]. Our model, complemented with high-temperature microscopic observations, can provide new atomistic insights into such transformations by identifying statistically important and previously unidentified structures in such complex grain boundaries. This can be further extended to two-dimensional materials where structurally diverse line defects periodically arrange at the nominal grain boundaries, controlling the growth kinetics and contributing to their electronic properties[48].

Our multi-scale hybrid model is motivated by the concept of emergence: that large and complex systems cannot be fully explained in a purely reductionist perspective [29]. Specifically, these systems can not be understood as simple extrapolations of reduced elements. As the system scales up, new models often emerge due to new physical constraints that become important only at longer length or time scales. In our case, a domain boundary is more than a random composition of lower-level structural motifs. At shorter length scales, our generative self-avoiding walker (Markov model) adheres to local motif-motif rules. At longer length scales, the domain boundary configurations are influenced by competing growth between epitaxial domains of two different possible configurations. These two models can readily hybridize (*e.g.*, Fig. 1) because they are interpretable.

Hierarchical assemblies of motifs emerge ubiquitously in nature, often due to selection pressures[49]. Similar hierarchies have been reported in colloids, nanoparticle assembly, supramolecular lattices, and more[10,11,25]. Such hierarchies usually point to emergent simplified descriptions, but they are often cloaked by randomness. While probabilistic models can reveal such emergence, these revelations can be lost in over-complex and uninterpretable models. In these cases, interpretable, generative models like ours can make sense of this randomness beyond what is practically observable, and have the potential to catalyze new directions for a common language to understand complex disordered materials[50–52].

## Methods

**Growth of potassium sodium niobate thin film**

$(K_xNa_{1-x})_yNbO_{3-z}$ epitaxial films were deposited on (001)-oriented 0.5% Nb-doped $SrTiO_3$ substrates by Radio Frequency (RF) magnetron sputtering, where x represents the relative fraction of potassium compared to all alkali content, y denotes the alkali-niobium ratio and z is the possible deviation in oxygen stoichiometry. Since K loss is more significant than Na, we used a sputtering target $(K_{0.55}Na_{0.50})NbO_3$ with excess content of K to compensate for the volatilization loss due to high temperature. By adjusting the target distance to the substrate, we can control the alkali-niobium ratio y. The sample under investigation in this work has a chemical formula of $(K_{0.26}Na_{0.74})_{0.52}NbO_{3-z}$. Detailed growth parameters have been reported somewhere else[34].

**STEM characterization and image simulation**

Atomic resolution STEM imaging was conducted using a JEOL ARM200F electron microscope equipped with a cold field emission gun, and an ASCOR 5th-order aberration corrector operating at 200 k$V$. HAADF images were acquired using inner and outer collection semi-angles of 68 and 280 mrad, respectively, with a convergence semi-angle of 30 mrad. ABF images were acquired using a collection semi-angle of 17 mrad. To improve the signal-to-noise ratio, a stack of 10 - 15 images was collected from a single region and then registered to an average image for both HAADF and ABF images. Multislice STEM imaging simulations were conducted by the *abtem* package[53] using an aberration-free probe with a probe size of approximately 1Å. The convergent semiangle was set to 30 milliradians (mrad), and the accelerating voltage was 200 kV in line with the experiments. HAADF images were simulated using a detector spanned an angular range of 68-280 mrad, and ABF images were simulated using a collection semi-angle of 17 mrad.

**Generation of motif planar graph**

The construction of the planar graph relies on the detection of graph nodes. We used the positions of Nb atomic columns as node positions, which are found by a maximum filter method and center-of-mass refinement[1]. For each node, we located all its natural neighbors using Voronoi tessellation and created edges from each node to its natural neighbors. This will result in a planar graph with most of the edges correct. However, it still suffers two main issues: overly lengthy edges near the graph boundary and overcounted neighbors too far away. The lengthy edge issue can be addressed by setting a maximum edge length, $L_{max}$, and any edges that exceed this length are removed. The overcounting issue can be resolved by introducing a weighting factor $s=L(f)/L$, $L(f)$ is the length of the Voronoi cell facet f separating the two neighboring regions that correspond to a given edge, and $L$ is the total peripheral of the Voronoi cell boundary. For each node, the weighting factor indicates the closeness of its natural neighbors. It should be noted that each edge can be seen by two nodes, thus giving two different weights ($s1, s2$). In our context, edges with $min\{s1, s2\} < 0.1$ are removed.

**Estimate the transition probability of the Markov model**

The transition matrix from motif state $i$ to motif state $j$, $P_{ij}$, is estimated by directly counting the number of occurrences of 3-motif configuration where state $i$ connects to state $j$. The transition matrix is then row-normalized. The ideal transition matrix is estimated in a similar way. Before row normalization, entries of the same color (*i.e.*, the same isometry group) are filled by their mean value.

### Estimate the unique counts of boundary segments

For a given boundary segment with $N$ motifs ($9 \leq N \leq 15$), we generate a pool of one million $N$-motifs obtained from our generative model. Next, we can estimate the unique count of equivalent boundary segments for each unique end-to-end vector. Lastly, we found that the average count $\Omega(N, R)$ has a power law relationship with $N$.

### Hybridize Markov model and random nucleation model

The translation symmetry of the domains allows two equivalent types of nucleation sites shown: type I (blue) and type II (orange). A Voronoi diagram can be constructed based on the nucleation sites. Domains of the same type will merge, leaving gray dashed edges. While edges shared by different domain types are highlighted by solid black segments, and they are coarse-grained trajectories of domain boundaries. These coarse-grained version boundaries can be decorated with sequential boundary segments generated from our Markov model. We randomly sample candidate segments from a pool of boundary segments to check whether their end-to-end orientations match the orientation of the coarse-grained boundaries. If matched version will be dressed along the coarse-grained domain boundaries.

### Acknowledgments

N.D.L. acknowledges funding support from the National Research Foundation (Competitive Research Programme grant number NRF-CRP16-2015-05), as well as the National University of Singapore Early Career Research Award. K.Y. and J.W. acknowledge the funding support from A*STAR, under RIE2020 AME Individual Research Grant (IRG) (Grant No.: A20E5c0086). J.W. also acknowledges the support by the Advanced Research and Technology Innovation Centre (ARTIC), through research project (ADT-RP2). This project is supported by the Eric and Wendy Schmidt AI in Science Postdoctoral Fellowship, a Schmidt Futures program.

### Contributions

J.D. and M.W. contributed equally to this work. J.D. conceived the idea, designed the framework, developed the algorithm, and conducted STEM simulations. M.W. grew the samples and conducted the STEM experiments. I.E. advised the domain boundary growth. S.J.P., K.Y., and J.W. advised on the experiments and the atomic models. S.J.P. and N.D.L. supervised the projects. J.D., S.J.P., and N.D.L. wrote the manuscript with input from all authors.

### Corresponding authors

Correspondence to Jiadong Dan and N. Duane Loh.

### Competing interests

The authors declare no competing interests.

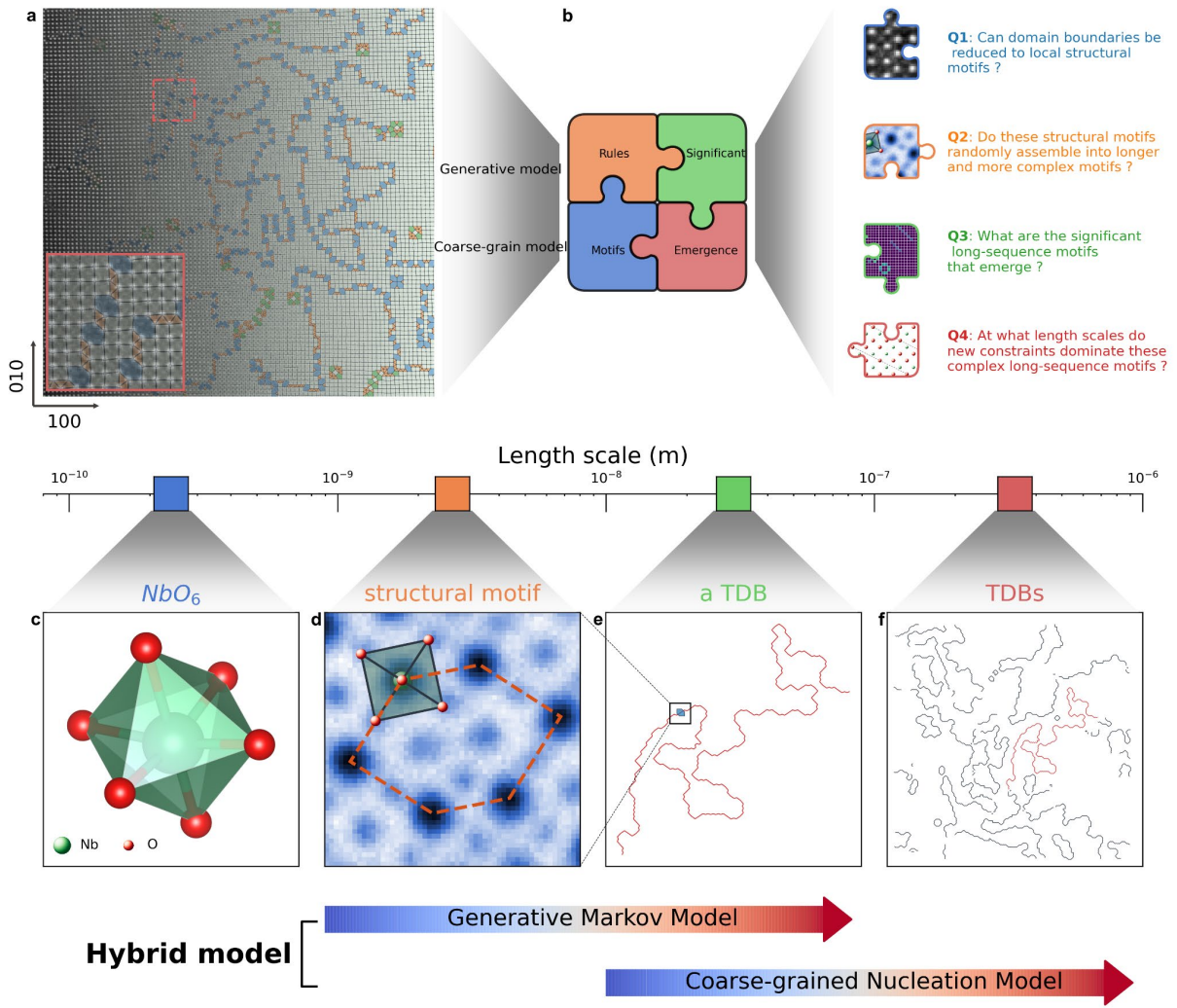

**Fig. 1 A hierarchical view of translation domain boundaries deciphered by a hybrid generative model.** The overall objective: **a** Decipher the motifs and their assembly rules to form seemingly disordered domain boundaries from their atomic resolution images. **b** This overall objective can be decomposed into four specific questions (Q1-Q4), which our hybrid model addresses. In a hierarchical perspective (from left to right, **c-f**), $NbO_6$ octahedra in **c** organize into structural motifs (orange dashed lines) along the domain boundaries in **d**, which in turn form a single domain boundary (TDB) in **e**; these TDBs compete as translational domains and pack densely on the plane.

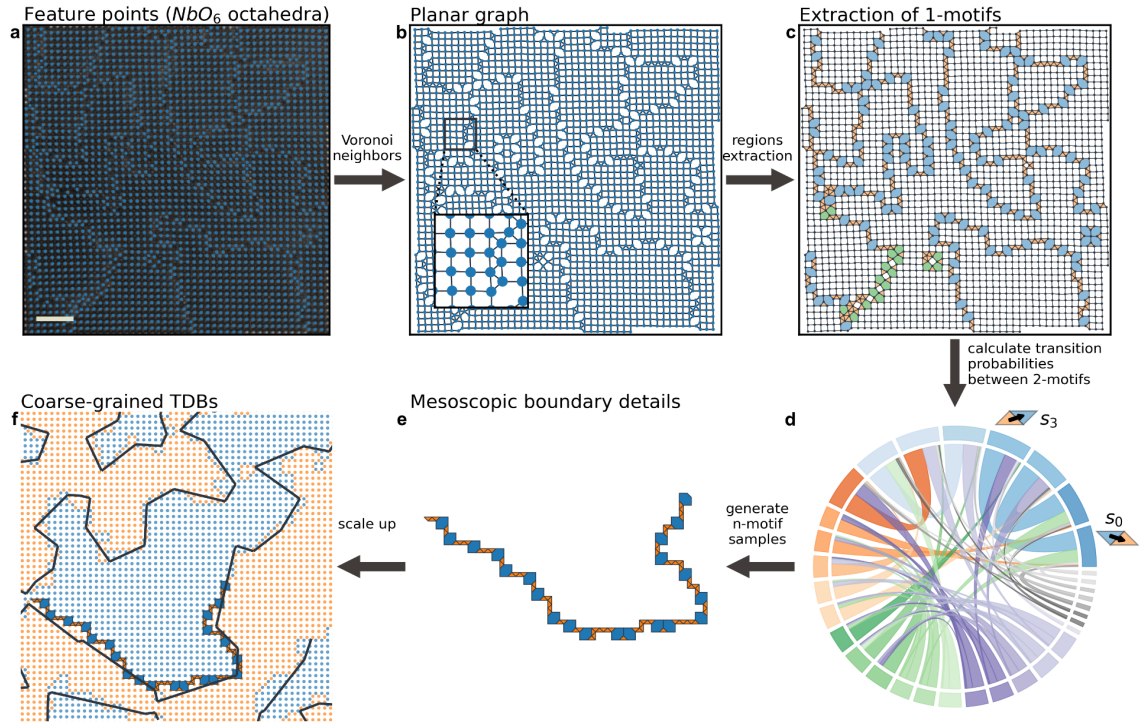

**Fig. 2 A workflow to efficiently represent structural motifs and recast the motif-motif assembly rules as a probabilistic generative model**. **a** A HAADF-STEM image of potassium sodium niobate (KNN) thin film abundant with translation domain boundaries (scalebar: 2nm). **b** Representing the KNN sample with a planar graph: graph nodes (vertices) are the Nb atomic columns in panel **a**; graph edges are constructed using Voronoi tessellation. **c** Programmatic extraction of 1-motifs (*i.e.*, polygons) from the graph in panel **b** and colored according to its number of sides. **d** Triangular and hexagonal 1-motifs form thirty-two 2-motifs on the observed domain boundaries, which can be described by the transition probabilities of a generative Markov model illustrated here as a chord diagram. **e** A synthetic domain boundary sample generated from our generative Markov model. **f** Replace coarse-grained boundary segments (solid lines) with randomly generated segments of motifs.

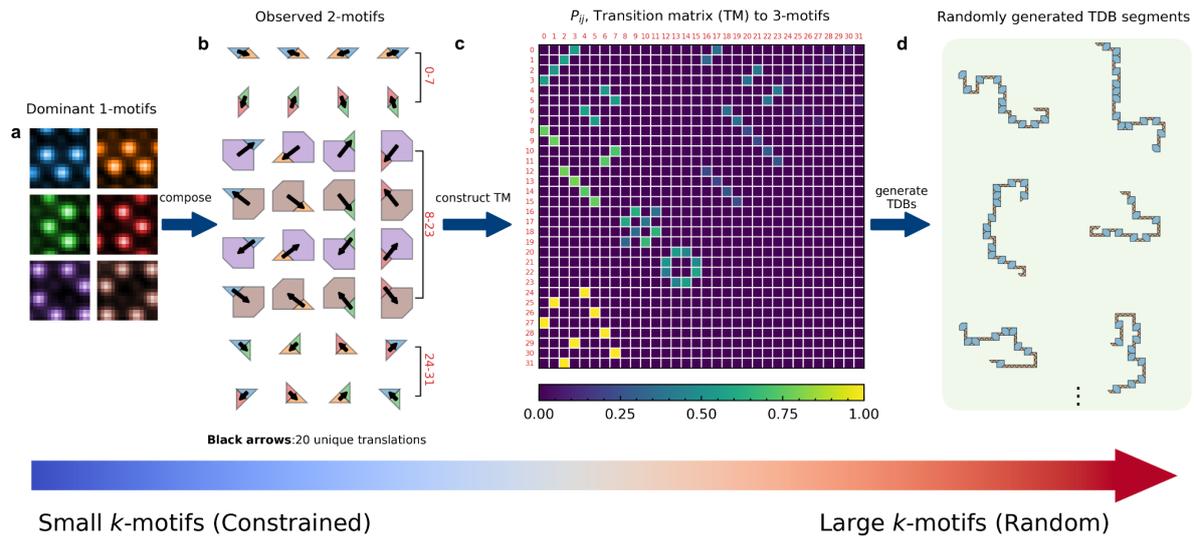

**Fig. 3 How the constrained *k*-Motif hierarchy leads to a probabilistic generative model**. **a** The six dominant 1-motifs identified and averaged from ADF-STEM images. There are 4 orientations for triangular 1-motifs (top) and two orientations for hexagonal 1-motifs (bottom). **b** The 32 unique observed 2-motifs that can be formed by the six 1-motifs. Based on Euclidean isometry (rigid transformation), they can be divided into three groups: 0-7, 8-23, and 24-31. **c** The transition matrix is estimated from our experimental data, and $P_{ij}$ is the probability of 2-motif *i* connecting to 2-motif *j*, where $0 \leq i \leq 31$ and $0 \leq j \leq 31$. **d** A collection of synthetic domain boundary samples generated from our model. From left to right, it shows a hierarchical structure of forming random complex domain boundaries from low-level constrained structural motifs.

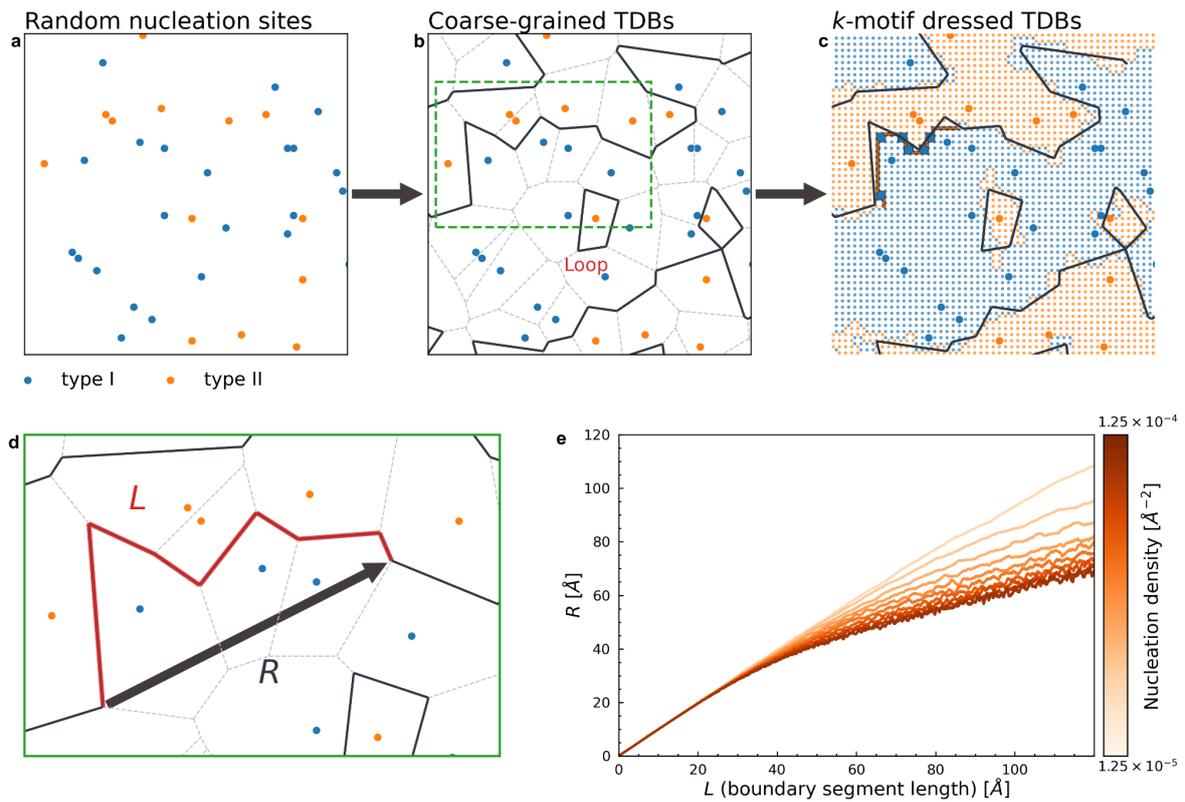

**Fig 4 Generating coarse-grained domain boundaries using a random nucleation model. a** Nucleation sites of two types of translation domains (*i.e.*, I, II) are randomly distributed across the substrate. Blue dots belong to the lattice from type I and orange dots are from type II. **b** Voronoi tessellation of nucleation sites in panel a; solid edges are the coarse-grained TDBs between different domain types while dashed edges are those between the same type. **c** We dress the coarse-grained TDBs in panel b with a possible arrangement of I, II translation domains. Here the edges in the latter are replaced by *k*-motifs generated from the Markovian Transition matrix. **d** The path length and end-to-end distance of a TDB segment are labeled here as e *L* and *R* respectively. **e** The relationship between *L* vs. *R* with increasing density of nucleation sites.

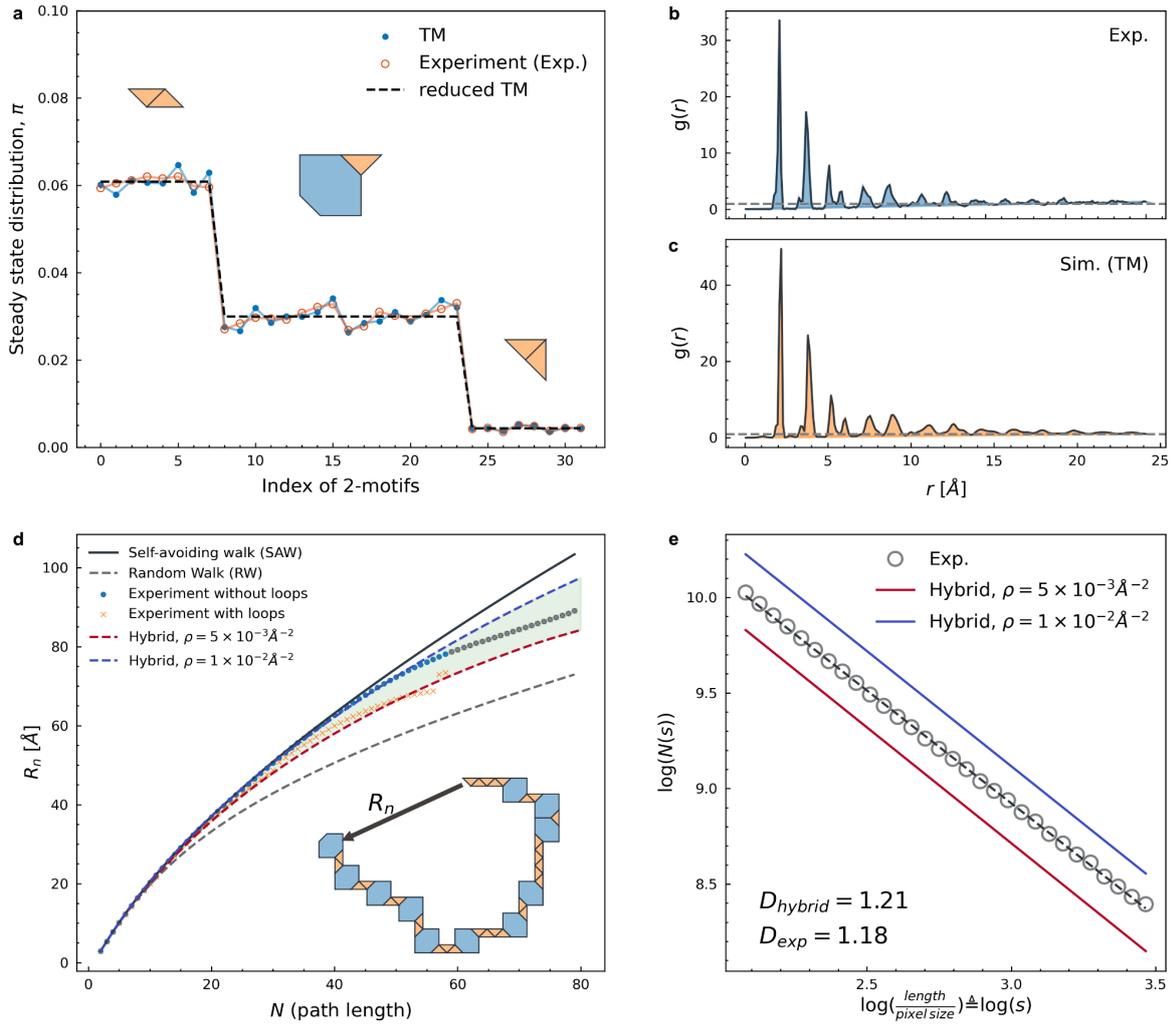

**Fig 5 Model evaluation and validation. a** The frequency distribution of thirty-two 2-motif states. This distribution shows three plateaus, where above each plateau we show its characteristic 2-motif (up to isometries). **b** The radial distribution function of all 1-motifs from experiments. **c** The radial distribution function of all 1-motifs from synthetic samples drawn using our hybrid model. **d** Comparison of end-to-end distance, $R_n$ (exemplified by inset), as a function of motif path length $N$ in different cases. **e** Fractal dimension of the domain boundary assemblies from the experiments (blue circles) and synthetic samples (orange lines). Both experiments and simulations show a power-law scaling with a fractal dimension of ~1.2.

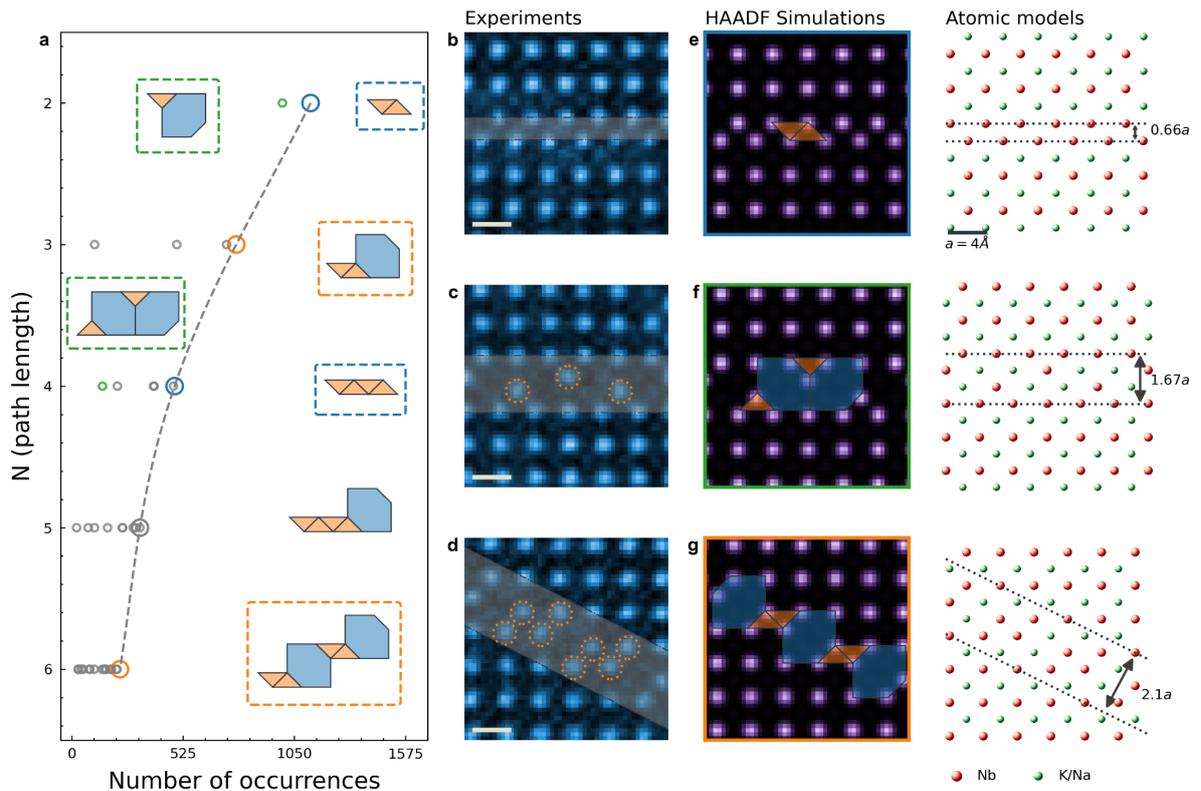

**Fig. 6 Significant tileable long sequences of *k*-motifs. a** Number of occurrences of tileable *k*-motifs that are statistically significant in Fig. 1a. Besides a simple chain of triangle-only motifs (blue dashed boxes), we observed a significant number of previously neglected longer sequences that contain hexagonal motifs (green or orange dashed boxes). **b-d** HAADF-STEM images showing single examples of the most frequently occurring, tileable, longer sequence motifs. **e-f** Multislice simulations of structures that match those in b-d respectively. **h-j** Atomic models of regions in b-d respectively.

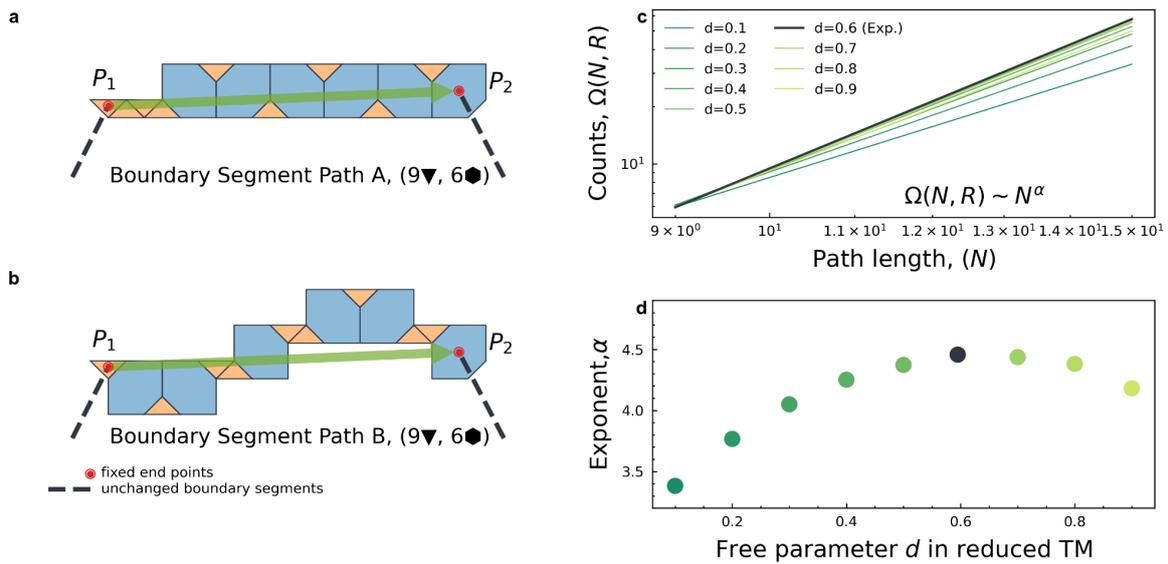

**Fig. 7 Estimating configurational entropy of domain boundaries. a, b** Two boundary segments that begin and end at equivalent fixed-end positions (red circles). Both domain boundaries have 9 triangular motifs and 6 hexagonal motifs. **c.** The average number of configurational microstates of boundary segments, $\Omega(N, R)$, increases with the segments' path length as a power law, shown for different values of the free parameter $d$ in the reduced transition matrix. **d**. The system maximizes the power law exponent $\alpha$ when the free parameter $d \approx 0.6$ (dark gray circle), which was observed experimentally.

Supporting Information

# A multiscale generative model to understand disorder in domain boundaries

**Authors:** Jiadong Dan[1,2,*], Moaz Waqar[3,4], Ivan Erofeev[1,5], Kui Yao[3], John Wang[3,4], Stephen J. Pennycook[6,7], N. Duane Loh[1,2,5,*]

**Affiliations:**

[1]NUS Centre for Bioimaging Sciences, National University of Singapore, 14 Science Drive 4, 117557, Singapore.

[2]Department of Biological Sciences, National University of Singapore, 16 Science Drive 4, Singapore, 117558, Singapore.

[3]Institute of Materials Research and Engineering (IMRE), A*STAR (Agency for Science, Technology and Research), Singapore 138634, Singapore.

[4]Department of Materials Science and Engineering, National University of Singapore, Singapore, 117574, Singapore.

[5]Department of Physics, National University of Singapore, 2 Science Drive 3, Singapore, 117551, Singapore.

[6]Department of Materials Science and Engineering, University of Tennessee, Knoxville, TN, USA.

[7]School of Physical Sciences and CAS Key Laboratory of Vacuum Sciences, University of Chinese Academy of Sciences, Beijing, China.

**Keywords:** scanning transmission electron microscopy, domain boundary, motif-hierarchy, generative model, coarse-graining

## Supplementary Section I: Derivation of steady-state distribution from the reduced TM and free parameter *d*

The transition matrix estimated from experimental data is sparse and block-like, which indicates that the parameter space of the transition probability matrix can be further reduced by symmetry. The sparsity stems from the fact that composing 2-motifs is not arbitrary and has to follow spatial constraints. For example, connecting 2-motifs indexed by 8 and 0 is allowed and gives a 3-motif as displayed in Fig. S6a. Composition of 2-motifs indexed by 8 and 1, however, is prohibited (Fig. S6b). After examining every index-pair (i, j), we have a total of 88 permissible 3-motif configurations represented by semi-transparent color blocks in Fig. S6c, where the light green entries are invalid and filled with zeros. These 88 permissible 3-motifs can be further classified into 11 classes because rotation and reflection are isometries (rigid transformation). In terms of model parametrization, the ideal system is fully determined by {a, b, c, d, f, g, k} as shown in Fig. S6d and tabulated in Table. S1.

Reduced by isometries, the steady-state distribution of 2-motifs should take the following form:
$$\pi = [x, x, \cdots, x, y, y, \cdots, y, z, z, \cdots, z].$$

There are 8 consecutive x, 16 consecutive y, and 8 consecutive z, therefore $\pi$ is a row vector of a length 32. If the transition probability is P, then the following relation holds,

$$P\pi = \pi.$$

We can obtain a system of linear equations as follow,

$$ax + cy + fz = x,$$
$$bx + (1 - c - k)y + gz = y,$$
$$8x + 16y + 8z = 1,$$
$$y(1 - d) + yd = y,$$
$$(1 - a - b)x + ky + (1 - f - g)z = z.$$

By solving this linear equation system, we can obtain x, y and z.

$$x = \frac{cf + cg + fk}{8(-ac - 2ag - ak - bc + 2bf + cf + cg + c + fk + 2g + k)}$$

$$y = \frac{-ag + bf + g}{8(-ac - 2ag - ak - bc + 2bf + cf + cg + c + fk + 2g + k)}$$

$$z = \frac{-ac - ak - bc + c + k}{8(-ac - 2ag - ak - bc + 2bf + cf + cg + c + fk + 2g + k)}$$

From our observations, k=0, g=0, and f=1, they can further simplify to forms as follow,

$$x = \frac{c}{16b + 16c - 8ac - 8bc},$$
$$y = \frac{b}{16b + 16c - 8ac - 8bc},$$
$$z = \frac{c(1 - a - b)}{16b + 16c - 8ac - 8bc}.$$

Notice that the parameter $d$ do not appear in these last equations, which indicates that $d$ is unconstrained by the 2-motif frequencies (*i.e.*, $x$, $y$, $z$), hence $d$ is a free parameter.

The specific values of {a, b, c, d, k, f, g} estimated from 2-motif and 3-motif counts are listed in Table. S1

| Parameters | a | b | c | d | k | f | g |
|---|---|---|---|---|---|---|---|
| Values | 0.555 | 0.374 | 0.764 | 0.594 | 0.0 | 1.0 | 0.0 |

**Table. S1 Free parameters for the reduced transition matrix (TM).**

**Supplementary Section II: Partial occupancy view of the grain boundaries**

Besides the k-motif view of the grain boundary structure, the periodic long sequence of grain boundaries can also be illustrated as partial occupancy of Nb octahedra in the gap between two grains.

If tiled periodically, the alternating structures of hexagonal and triangular units in Fig. 6c can be regarded as extra Nb octahedra (dashed orange circles) partially occupying certain positions in the gap (half-transparent band) of the upper domain and half-translated lower domain. In terms of atomic structure models, it corresponds to atomic model Fig. S10a, where the extra Nb octahedra are highlighted in light purple color. In Fig. S10a, the alternating periodicity of the purple octahedra attaching to the left and right domains is 1. Increasing this periodicity into 2 and 3, we can obtain new grain boundary configurations in Fig. S10b and S10c, which are less frequently observed long sequences in our experiments. In particular, as illustrated in Fig. S10d, the most commonly observed configuration is comprised of all triangle motifs and is also compatible with the occupancy model: all the extra octahedra occupy either side of the domains. The slanted grain boundary (Fig. 6d) has a wider gap (Fig. 6j), and the extra Nb octahedra along with octahedra from two grains are arranged into parallelogram units to occupy the gap.

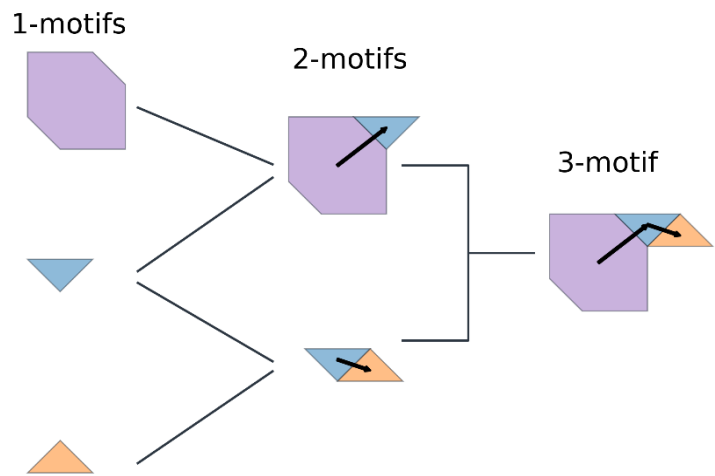

**Fig S1. Polygonal representation of 1-motif, 2-motif, and 3-motif in a hierarchical way.** We define individual polygons here as 1-motif (mono-motif). Two connecting 1-motifs can combine to form a single 2-motif; Two 2-motifs can organize themselves to obtain a 3-motif configuration. If the process continues, by definition, k-motifs are connected motif sequences of path length k.

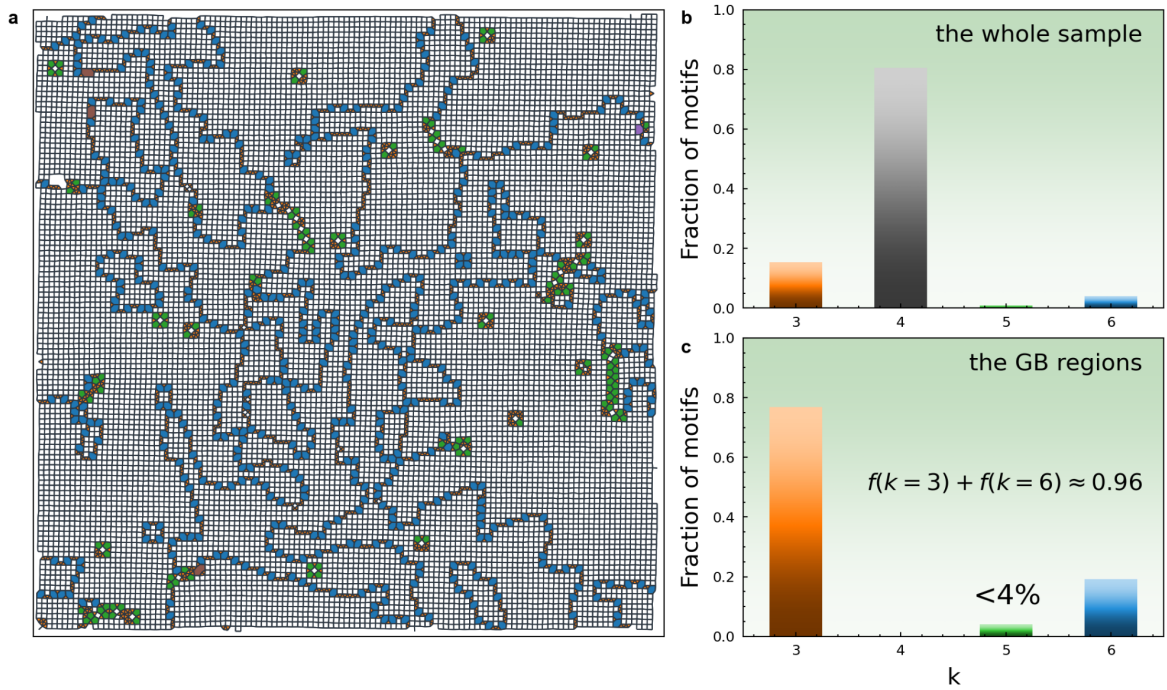

**Fig. S2 The ring statistics of 1-motifs. (a)** Programmatic extraction of 1-motifs (*i.e.*, polygons) from the KNN sample. **(b)** Histogram of 1-motifs from the field of view in panel **a**. **(c)** Histogram of 1-motifs from the grain boundary region in panel **a**. The pentagon motifs only account for less than 4%.

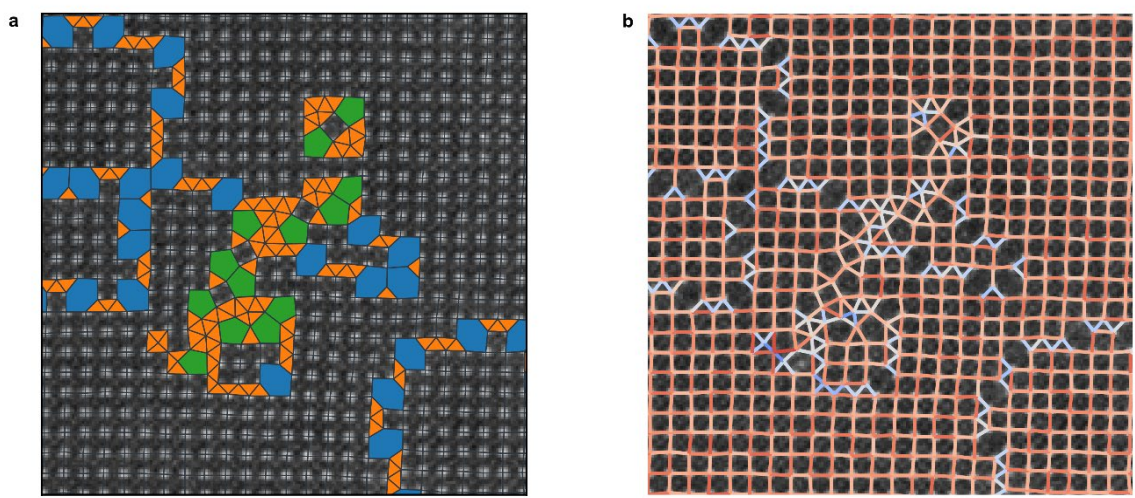

**Fig. S3 Frustrated regions have pentagonal motifs. (a)** The selected region shows the presence of pentagonal motifs. **(b)** A map showing the in-plane distances between two Nb atomic columns.

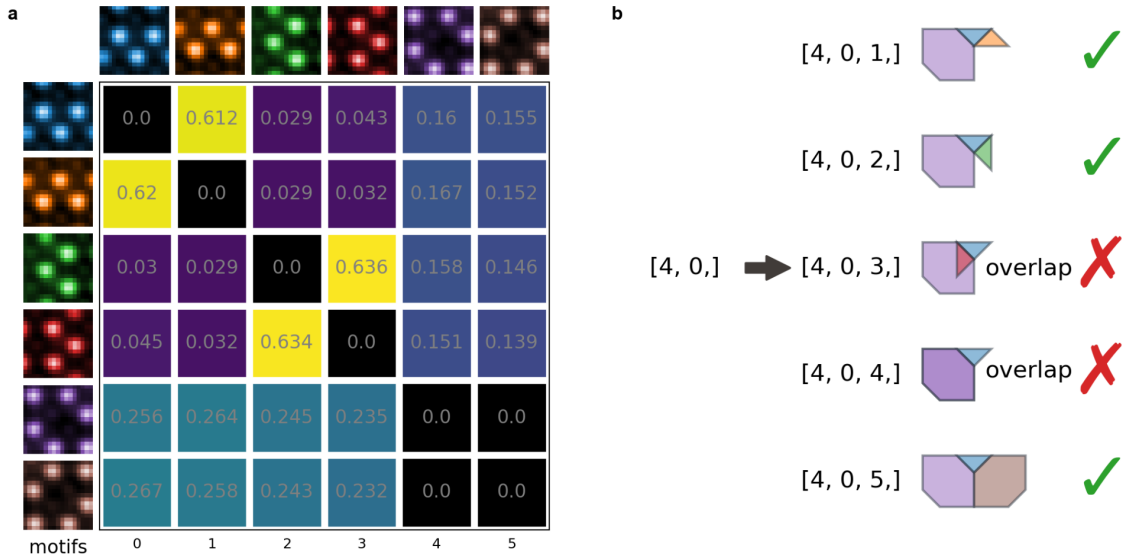

**Fig. S4 Using 1-motifs as Markov states can generate unrealistic samples. (a)** Left panel: Using the six dominant 1-motifs as motif states in the Markov model, we can obtain a transition matrix that combines two 1-motifs to form a 2-motif. **(b)** The notation [4, 0] means motif 4 (purple hexagon) is followed by motif 0 (blue triangle). According to our transition matrix, motif 0 can be followed by {1, 2, 3, 4, 5}, thus producing five samples represented by [4, 0, i] (1≤i≤5). Among these five samples, [4, 0, 3] and [4, 0, 4] are not realistic since the third motif overlaps with the first motif. To obey the spatial constraints, the direction of neighboring 1-motif centers should be incorporated into our model.

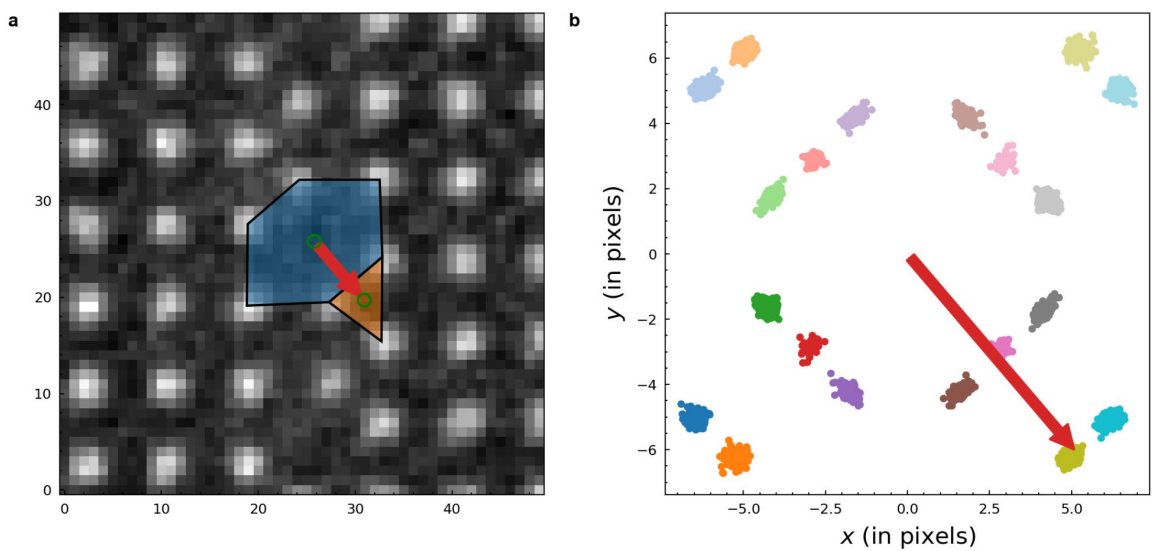

**Fig. S5 A total of 20 classes of translation vectors associated with 2-motifs.** The left panel illustrates that the translation vector (red arrow) is started from the center of the initial 1-motif to the center of the end 1-motif. The right panel visualizes all translation vectors in our experiments. There are 20 fuzzy clusters, also indicating the sample is strained to some degree.

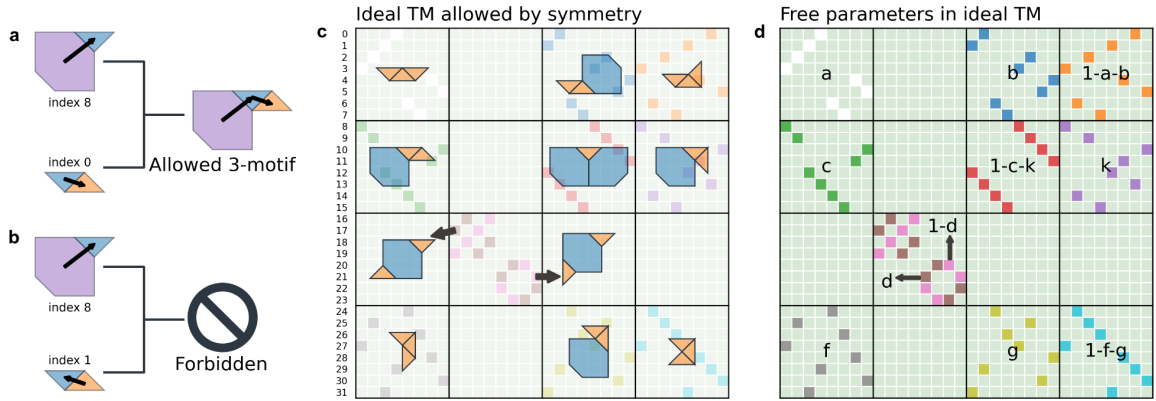

**Fig. S6 The ideal generative model reduced by the isometries of 3-motifs.** The transition matrix that encodes the local rules is sparse since composing 2-motifs is not arbitrary. For example, panel **a** shows 2-motifs indexed by 8 and 0 can combine to form an allowed 3-motif, while panel **b** illustrates the combination of 2-motifs indexed by 8 and 1 is prohibited. In total, there are 88 permissible 3-motifs, which means the transition matrix has 88 valid inputs. These 88 permissible 3-motifs can be further classified into 11 classes because rotation and reflection are isometries. **c** 11 Unique color blocks in the ideal TM mark the resultant 3-motifs overlaid on top. **d** In terms of model parametrization, the ideal system is fully determined by {a, b, c, k, d, f, g}.

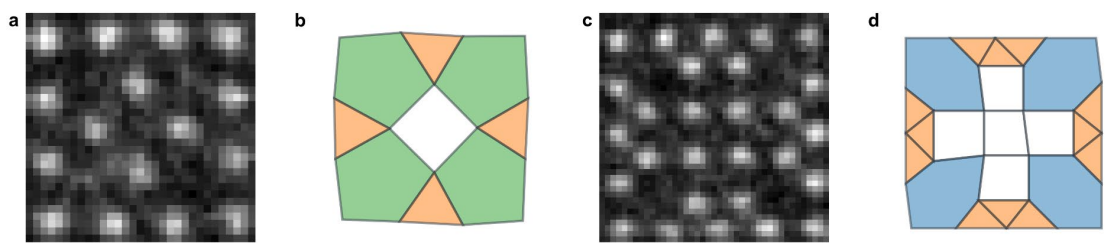

**Fig. S7 Two typical small loop structures.** (**a**) A small square-like loop comprising triangular and pentagonal 1-motifs. (**b**) motif representation of the loop structure in panel **a**. (**c**) A small square-like loop comprising triangular and pentagonal 1-motifs. (**d**) motif representation of the loop structure in panel **c**.

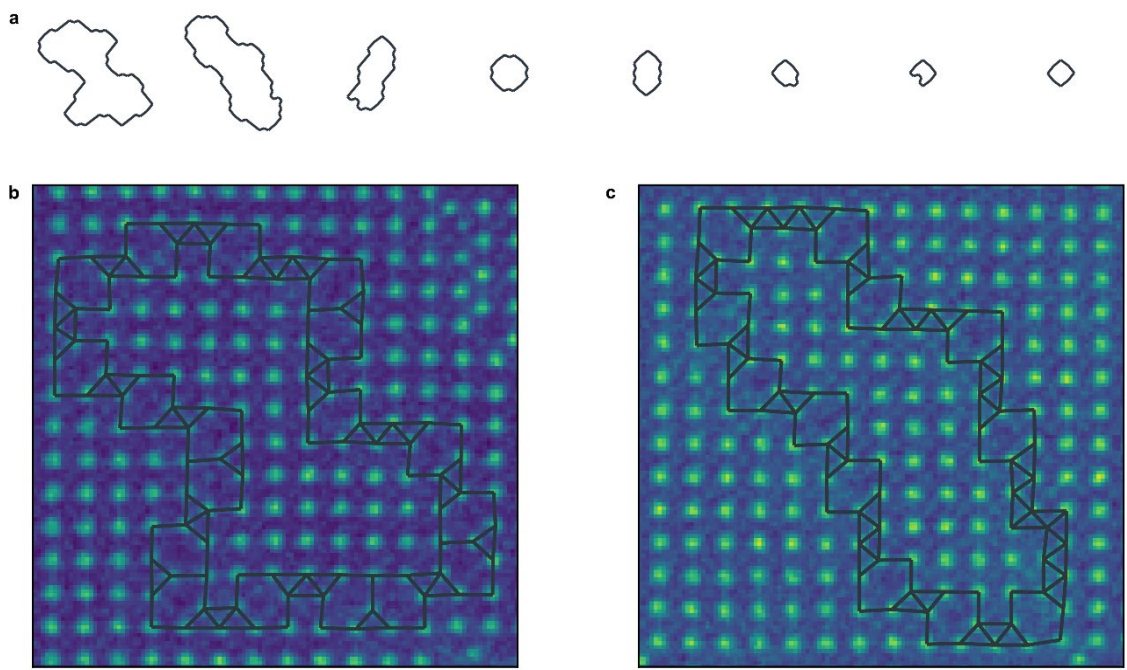

**Fig. S8 Selected larger loop structures comprising triangle and hexagon 1-motifs. (a)** A set of loops structures represented by edge segments. **(b)** An ADF-STEM image shows the first loop structure in panel **a**. **(c)** An ADF-STEM image shows the first loop structure in panel **a**.

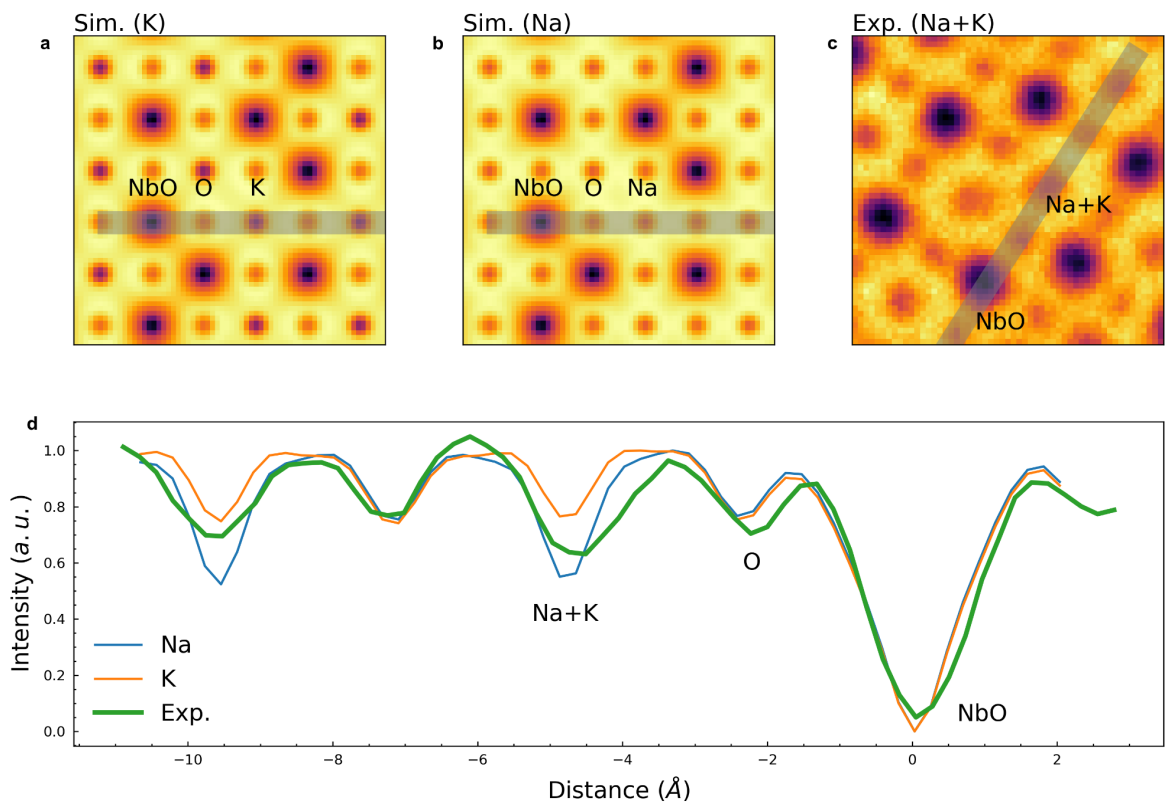

**Fig. S9 Chemical information inside the hexagonal 1-motif.** **(a)** A simulated ABF-STEM image of hexagonal 1-motif, which only contains potassium and oxygen atomic columns. **(b)** A simulated ABF-STEM image of hexagonal 1-motif, which only contains sodium and oxygen atomic columns. **(c)** A mean patch of the hexagonal 1-motif region extracted from experimental ABF-STEM images. **(d)** The line profile analysis from images in panels a, b, and c. Panel d shows that the atomic columns inside contain a mixture of potassium and sodium.

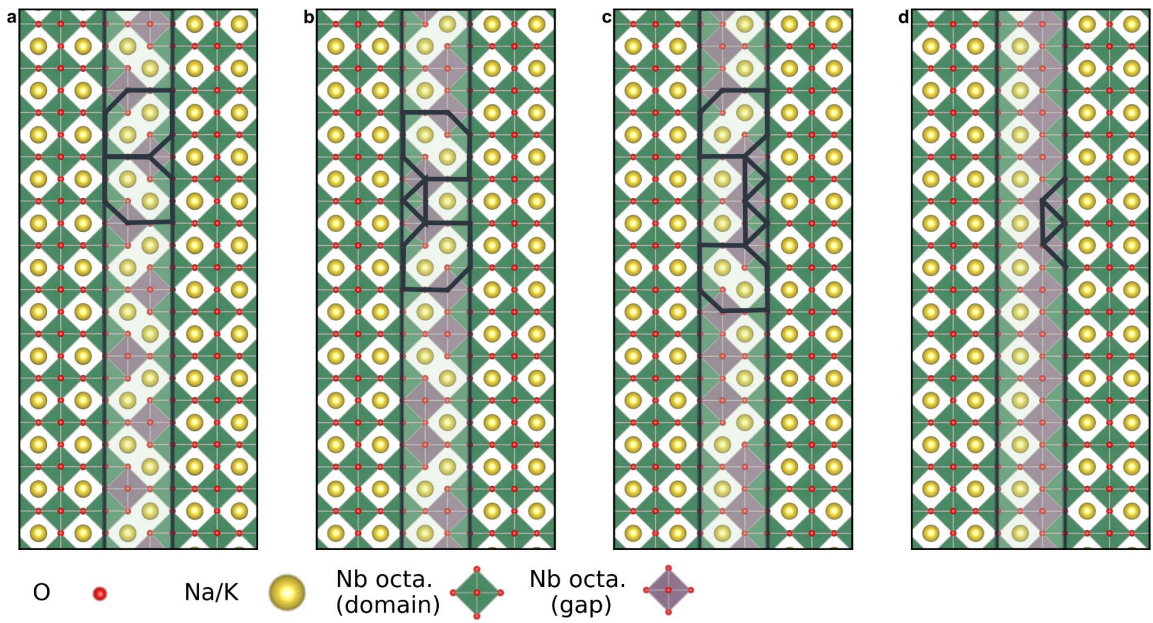

**Fig. S10 Different possible periodic boundaries between two translated domains.** Atoms on the left (or right) of the region between the solid black lines in all four panels are identical and belong to the same translation domain. However, there are still many possible atomic configurations in the grain boundary straddled between the black lines. Panels **(a-d)** show only four of such possible configurations. The Nb atoms, which are coordinated between six oxygen atoms (red spheres), are shown as either green or purple octahedra where the latter lie within the grain boundary.

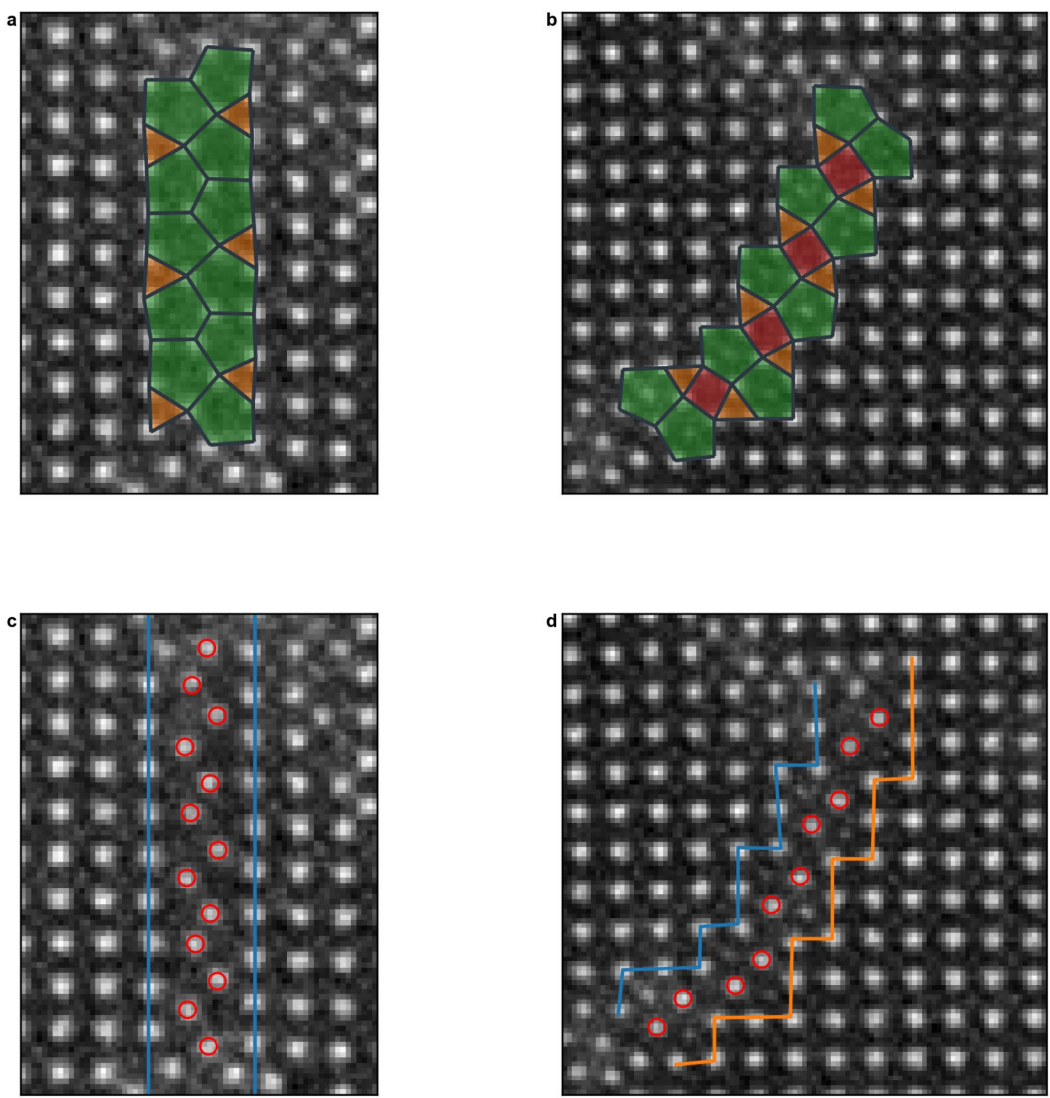

**Fig. S11: Experimentally observed long-sequence motifs comprising pentagonal motifs. (a)** Pentagonal and triangular motifs can assemble into a short-range ordered linear structure. **(b)** Pentagonal, square and triangular motifs can assemble into ordered structures. **(c)** The partial occupancy view of the ordered structure in panel a. **(d)** The partial occupancy view of the ordered structure in panel b. The blue and orange lines indicate the edges of two misaligned domains. In the lower row of panels, the red circles highlight the positions of inserted $NbO_6$ octahedra.